\documentclass[submission,copyright]{eptcs}
\providecommand{\event}{F-IDE 2014} 
\usepackage{graphicx}
\usepackage{enumitem}
\usepackage{wrapfig}

\title{SPEEDY: An Eclipse-based IDE for invariant inference}
\author{
David R. Cok
\institute{GrammaTech, Inc.\\
Ithaca, NY, USA}
\email{cok@frontiernet.net}
\and
Scott C. Johnson
\institute{GrammaTech, Inc.\\
Ithaca, NY, USA}
\email{sjohnson@grammatech.com}
}

\def\titlerunning{SPEEDY}
\def\authorrunning{Cok \& Johnson}
\def\copyrightholders{2014. GrammaTech, Inc.}
\begin{document}
\maketitle

\begin{abstract}
SPEEDY is an Eclipse-based IDE for exploring techniques that assist users in
generating correct specifications, particularly including invariant inference algorithms and tools.
It integrates with several back-end tools that propose invariants and will incorporate 
published algorithms for inferring object and loop invariants. Though the architecture is language-neutral, current SPEEDY targets C programs. Building and using SPEEDY has 
confirmed earlier experience demonstrating the importance of 
showing and editing specifications in the IDEs that developers customarily use, 
automating as much of the production and checking of specifications as possible, 
and showing counterexample information directly in the source code editing environment. 
As in previous work, automation of {\em specification checking} is provided by back-end SMT solvers.
However, reducing the effort demanded of software developers using formal methods also requires
a GUI design that {\em guides} users in writing, reviewing, and correcting specifications and
automates {\em specification inference}.
\end{abstract}

\section{Introduction}

\subsection{GUIs for software verification}

While theoretical work on software verification technology has proceeded rapidly in recent years, there has not been a corresponding increase in using software verification for day-to-day software development. Where applied to industrial software, formal methods have remained in the hands of experts in logic and formal software verification.

A contributing cause of this problem is the lack of tools and automation that put verification tools  where typical software developers need it - directly within the IDE used for other software development tasks. Furthermore, for good usability, the formal methods detail is best hidden behind a screen of automation: a work-flow that requires off-line processes or pops up windows containing interactive provers and logically stated verification tasks will not, in the authors' opinion, gain wide acceptance by developers already leery of static analysis. 

We posit that the best IDEs incorporating formal methods technology must automate as many tasks as possible, including both specification discovery and specification checking; it must also present specifications and counterexample information directly in the user's working environment; where automation is not possible, specification templates and helpful advice to the user should be available. Stated 
differently, the GUI for software verification and the GUI for software development should be one and the same, with an integrated set of actions and features. Our vision is to create unified IDEs for
managing software verification tasks, thoroughly
integrated with commonly used software development tools, hiding and automating as much of the formal methods detail as possible. We see this task as combining design of user assistance features and  integrating and {\em reducing to practice} the 
variety of advances in verification technology over the past decade.   

This paper describes SPEEDY, a tool for specification editing, discovery, checking and debugging. Some of these elements have been demonstrated in previous work (e.g., \cite{Cok-2011-OpenJML}, \cite{Cok-2011-jSMTLIB} and subsequent unpublished developments), but are further integrated here. SPEEDY's
particular contribution is to integrate techniques for (a) guided writing and review of specifications and (b) automated specification discovery (a.k.a. invariant inference, or function summarization)\footnote{We use the terms invariant inference, specification discovery, and function summarization interchangeably.} into the user's working environment. Furthermore, while most work in this vein is for object-oriented languages (e.g., Java, C\#, Dafny, Scala), SPEEDY targets C and C++.
SPEEDY supports a development style in which top-level specifications are generated from requirements, and then refined into code and detailed specifications developed in tandem, with automated support for checking the code and specifications.

The project is funded by NASA as SBIR contracted research. This report describes the results of the Phase I research and the goals of the next phase. Thus, the tool is a work in progress; Phase I is designed to resolve technical risk and demonstrate a proof of concept. The full implementation of the prototype is a task for Phase II.

\subsection{Humans and software verification}
\label{Review}

One strand of work in static analysis seeks to provide fully automated checking of software. This endeavor has seen significant advances in tools for model checking and abstract interpretation. However, the need for scalable, automated inter-procedural analysis has required approximations that limit the precision and sometimes the soundness of the resulting tools. Nevertheless, such tools are
successfully distributed commercially and applied to realistic software.

However, a fully automated solution, in which the user never writes or sees a specification, is limited: it can only check the {\em implicit specifications} of the target software. By implicit specifications we mean those restrictions imposed by the programming language semantics, such has not dividing by zero, not accessing an array out of bounds, and not dereferencing an invalid pointer. In contrast, explicit specifications are those that state what the program is supposed to do; only with
a complete, or at least substantial, set of (valid, accurate) explicit specifications can we be sure that a target program does what is intended. A useful 
software verification system will check that the target program is consistent with both the implicit specifications and any explicit specifications.

Where are the explicit specifications to come from? 
Traditionally such specifications are provided by domain experts and software verification experts, working together. This is typically a costly, labor-intensive endeavor. Thus the interest in
another strand of work in program analysis: {\em invariant inference}. This research attempts to 
infer invariants about a program from the program source code. In typical specification and verification systems, the user must write specifications for each module and invariants for each loop within modules. Having automated inference techniques at one's disposal will reduce, significantly, one hopes, the effort needed to verify a set of software. One goal of the SPEEDY project is to integrate a variety of such 
techniques into an IDE for software development and verification; success at this integration will be a significant advance in usability of verification systems.

However, humans must stay in the software verification loop for two reasons.
First, invariant inference algorithms, operating on source code, can only infer what existing software 
actually does, not what it is supposed to do. Thus humans are still needed to review the products of inference algorithms to check that automatically generated invariants indeed state the intent of the software. Nevertheless, it is less work to validate, correct or generalize purported specifications than it is to write 
specifications from scratch. Even if only the simpler 50\% of specifications that are automatically generated, a substantial amount of thinking and writing by the user can be avoided. Importantly, because humans are still a part of the process, it is essential that the composite verification system have a design that promotes usability for all levels of expertise. 

Second, a software verification system can only verify that specifications and code are consistent, 
and not that both are correct. Careful human review processes are needed to be sure that informal
requirements are correct, complete, and accurately translated into the target specification language.

Thus, SPEEDY combines GUI features for effective user review and manipulation of specifications with techniques for automatically suggesting specifications.

\subsection{Contributions}

SPEEDY is built on the Eclipse CDT (Eclipse's C/C++ development environment), uses back-end SMT solvers, and integrates various invariant inference tools. 
It is a tool for specification discovery and review, integrated into a full software and specification development environment. As in previous tools, SPEEDY provides back-end checking of specifications using SMT solvers.
The key contributions of SPEEDY are these:
\begin{itemize}
\item it provides assistance to users in generating and reviewing specifications, building on programming productivity research by Schiller and Ernst \cite{SchillerE2012};
\item as one element of user assistance, SPEEDY integrates various invariant inference algorithms;
\item SPEEDY provides an IDE for debugging specifications integrated with the software development environment, hiding most aspects of provers and formal methods;
\item SPEEDY is built on the Eclipse CDT, providing GUI features that integrate specification 
manipulation with code development;
\item SPEEDY is constructed with an architecture that enables evaluation of alternatives in tools and design decisions.
\end{itemize}
 
The following sections describe the architecture and main elements of SPEEDY and our progress in implementing our overall vision.

\section{Architecture of SPEEDY}

The principal goal of the SPEEDY project is to integrate, study and evaluate invariant inference tools and algorithms. However, it is also desired to assess their use in the context of actual software development. Thus SPEEDY is built as a full specification authoring, editing, discovery, checking, and
debugging tool, fully integrated into an IDE for software development. 

We chose Eclipse as the IDE and the Eclipse CDT (C Development Toolkit) as the C/C++ software development environment for two reasons: it was a UI environment with which we were familiar, having generated other Eclipse plugins in the past, and the contract customer desired an interface that was familiar and freely distributable. However, the more significant actions all can be invoked as command-line actions and could as well be interfaced to an alternate IDE, such as emacs or Visual Studio.

\begin{figure}
\centering
\includegraphics[width=0.48\textwidth]{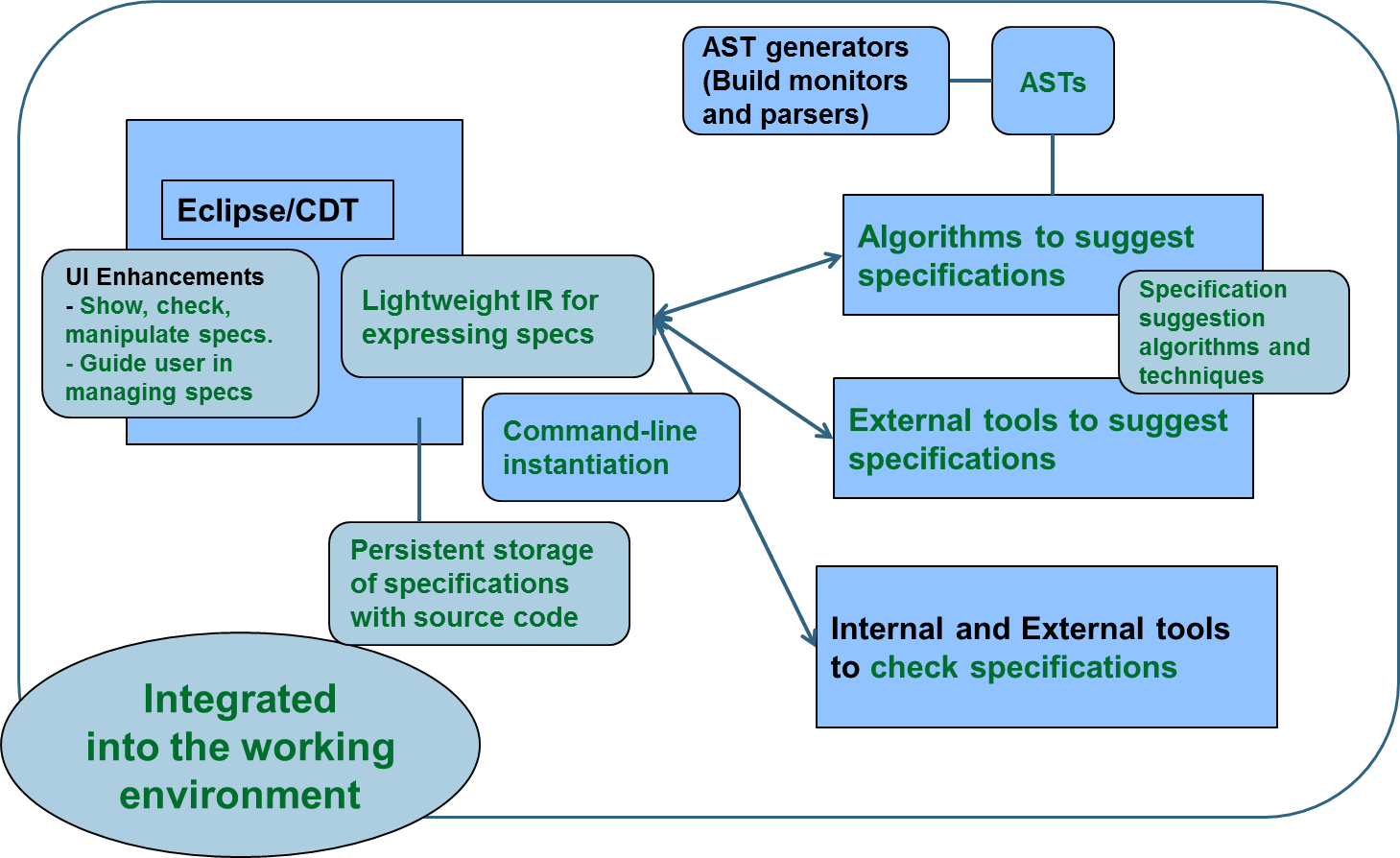}
\caption{The architecture of the SPEEDY system.}
\label{Fig:architecture}
\end{figure}

The architecture of the system is shown in Fig. \ref{Fig:architecture}. 
At the center of the architecture is a database holding internal representations of specifications in association with the corresponding source code. There are processes for reading and writing these from persistent storage; the specifications, if separate from the source code, can themselves be saved in the same version control system. Similarly the GUI interacts with the specification database to
display, edit, and otherwise manipulate specifications, using a conventional model-view-controller design pattern. 

The GUI itself provides mechanisms to guide users through specification writing and 
reviewing tasks, using Eclipse UI elements. The most significant of these is to integrate invariant
inference tools and algorithms. Those also interface with the specification database through a common API, making it easy to add new inference tools and algorithms. The inference algorithms make use of 
sources of static analysis information in the form of abstract syntax trees attributed with type and symbol information. 

Finally SPEEDY uses SMT solvers to check specifications, receiving counterexample information from the solver; the counterexample information is then displayed to the user in the editor window for the corresponding source code.

An important element of the architecture is that it be a platform for exploring various design decisions or adding new or replacement modules in the future. This is useful during the prototyping phase, but also as a research platform. Modular elements include these:
\begin{itemize}[noitemsep]
\item the storage format, which is currently XML but could be any other format (section \ref{Storage})
\item the specification display format - currently we use an ACSL \cite{ACSL} format and experimented with VCC \cite{VCC};  changing this is as easy as supplying a new parser and pretty printer (section \ref{Language})
\item the inference tools and techniques - we integrated five different approaches, with plans for more (section \ref{Inference})
\item sources of ASTs - four are described in section \ref{ASTs}
\item specification checkers - we experimented with both Frama-C \cite{webframac} and direct translation to SMT (section \ref{Debugging})
\item SMT solvers - we used CVC4 \cite{webcvc4}, Z3 \cite{DeMoura:2008:ZES:1792734.1792766}, and (through Frama-C) alt-ergo \cite{altergo-web} - integration with Why \cite{Why3} is planned for the future (section \ref{Debugging})
\end{itemize}

\section{GUI for manipulating specifications}

\subsection{Specification language}
\label{Language}

At the beginning of the project we expected to spend some time assessing  
specification languages, and needing to choose one on which to base SPEEDY.
SPEEDY needs a specification language in which it is possible to express, at least in principle, the full functionality of procedures, and not simply a relatively simple 
keyword language such as SAL or a C-equivalent of Java annotations. Just two languages were viable: ACSL (ANSI-C Specification Language) \cite{ACSL} and the VCC input language \cite{VCC}. The conclusion of this investigation was that 
these two are similar enough that a common internal representation can be used, with translation to
either language as needed for display or interaction with external tools.

\subsection{Storing and displaying specifications}
\label{Storage}

A key design decision in a system managing code and specifications is where to store the specifications.
For some languages, such as Java, the specifications can be naturally written directly in the .java
files. For C, interface specifications should be associated with .h files and specifications within
implementation code (such as loop invariants) will be in .c files.
However, there are situations in which the source files should not or cannot be modified. The source files may not be available, such as when a library interface is being specified. Or, the development
team may wish to keep the specifications separate from the source. 

Accordingly, there are common use cases in which the specifications are physically separate from the source code with which they are associated. Nevertheless, to facilitate easy and accurate editing of code and specifications, the UI should present the user with the specifications directly in conjunction with the code. We call this process {\em weaving}: code and specifications are combined 
together in the editor from separate sources, and then split apart again when saved. The goal is to provide a seamless display and editing experience.

\begin{figure}
\includegraphics[width=\textwidth]{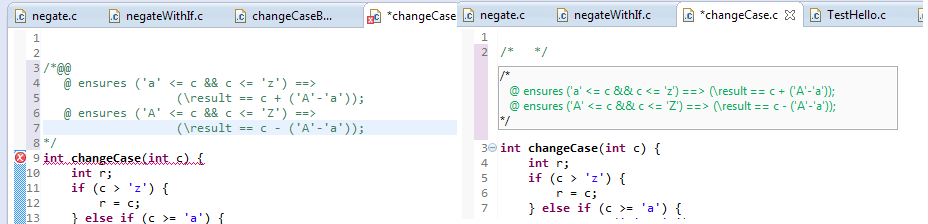}
\caption{Two ways to weave specifications with source}
\label{Fig:weaving}
\end{figure}

SPEEDY prototyped two ways of weaving source and specifications in the display, as shown in Figure \ref{Fig:weaving}. 
The first inserts the specifications as comments directly in the edit buffer, here matching the style of ACSL (though other
styles could readily be used). The advantage is that no special graphics are needed; the disadvantage
is that the line numbers change and adjustments have to be made to line numbers that reference the
original source file. The second style of weaving, on the right, shows the specifications in a special, editable, styled text box 
inserted into the edit buffer. The text box is set to take up vertical space corresponding to its contents;
thus this display style has the advantage of not changing the line numbering; the disadvantage is
that careful attention to refreshing the display is required to keep the textboxes refreshed and displayed in their
correct locations.

\subsection{Guiding the user in creating specifications}

A key element of SPEEDY is implementing mechanisms to guide the user in creating specifications. In this we are inspired by the work of Schiller and Ernst \cite{SchillerE2012}. In their VeriWeb system, they measured the
effects of various kinds of guidance and advice to users on the users' ability to generate and check specifications.
Table \ref{Table:GUI} shows the advice mechanisms proposed by Schiller and Ernst and the corresponding
implementation, with modifications and extensions, in SPEEDY.

These UI elements have two key goals. First, some of them are meant to speed up editing by incorporating mouse or keyboard actions that accomplish routine tasks. More importantly, the user's macro-level tasks and understanding are facilitated by offering automated advice and suggestions for 
new specifications or to correct specification errors.
\begin{table}
\begin{tabular}{|p{.47\textwidth}|p{.47\textwidth}|}
\hline
{\bf Schiller and Ernst research result} & {\bf SPEEDY implementation} \\
\hline
Drag and drop editing interface  & 
Standard Eclipse editing features: templates, code completion, context assist, keyboard shortcuts \\
\hline
Concrete counterexamples (from execution traces) &
Concrete counterexample values and paths, from static analysis, overlaid on the source code implementation \\
\hline
Specification inlining, in the web tool &
Two modes of specification inlining, in the source code editor\\
\hline
Context clues &
Suggested specification locations \\
\hline
Specification suggestions from Daikon &
Specification suggestions from multiple tools, including Daikon \\
\hline
Active guidance &
Active guidance using integrated Eclipse Wizards, template-guided editing, and quick fixes to give suggestions \\
\hline
\end{tabular}
\caption{GUI elements implemented in SPEEDY}
\label{Table:GUI}
\end{table}

\subsection{Other GUI elements}

SPEEDY made full use of the functionality available in Eclipse, implementing many other features that
are (fairly straightforwardly) available through Eclipse extensions:
\begin{itemize}[noitemsep,nolistsep]
\item tracking the provenance of specifications (user written or tool generated)
\item a custom View to show status of specifications: Auto-generated but not yet reviewed, Accepted, Rejected, Valid, Invalid
\item syntax highlighting to show status
\item integration with the Eclipse build system (to launch static analysis tools)
\item a custom view to show and edit specifications (as well as showing them in the editor)
\item Eclipse problem markers to identify syntactic problems or failed specification checks
\item showing counterexample information from failed proof attempts directly in the C editor windows, in terms of the source code (cf. section \ref{Debugging})
\item Eclipse markers to identify locations where specifications should be inserted
\item Quick fixes to insert inferred specifications in appropriate code locations
\item code completion and content assists, to streamline editing
\item menu items to launch various actions, including the ability to map key combinations to actions
\item Eclipse preference and property sheets
\item Eclipse perspective for manipulating and generating specifications
\item packaging as a standard Eclipse plugin
\end{itemize}

\noindent A few more complex items were left for Phase II of the prototype development:
\begin{itemize}[noitemsep,nolistsep]
\item Integrating the parsing (and syntax error checking) of specifications with that of code
\item Syntax highlighting of keywords in specifications
\item Implementing Wizards as a way to guide users through the steps of writing specifications
\item Integrating specifications into the CDT's refactoring and searching tools.
\end{itemize}

Two of the above items are worth elaboration. First, SPEEDY tracks the status of specifications. As we discussed in section \ref{Review}, a specification generated automatically may need human review -- to accept it as is, to correct, to generalize, or to discard. By indicating a status -- user-written, user-accepted, user-rejected, or not yet reviewed -- progress of work can be recorded and remembered, even on a large scale or in a team environment. Also, changes in generated specifications or out of date specifications or checks can be highlighted.

Second, we expect to use Eclipse Wizards as a way of walking the user through the process of writing or
reviewing specifications. For example, a method needs pre- and post- and frame-conditions; it is helpful to be reminded of the various pieces. Similarly loop conditions have a number of interacting aspects that are easy to omit, if not reminded of each one.

\section{GUI for checking and debugging specifications}
\label{Debugging}

Specifications will never\footnote{well, hardly ever} be correct without automated means to check them; furthermore, just as with code, effective debuggers are needed to evolve specifications and code
into mutual consistency.

SPEEDY uses a now common paradigm for modular checking of code and specifications (for example, see \cite{Cok-2011-OpenJML,Leino05,FlanaganSaxe01,Barnett-Leino05}). Checking is performed procedure by procedure:
\begin{itemize}
\item the procedure specifications are translated into assumptions and assertions interleaved with the code, based on the semantics of the specification language. For example, preconditions become assumptions at the beginning of the procedure and postconditions are assertions at the end.
\item the code, assumptions, and assertions are translated into a basic block form that uses 
single-assignment labeling of variables
\item the basic blocks are translated into compact verification conditions (VCs)
\item the verification conditions are expressed in SMTLIBv2 \cite{BarST-RR-10,BarST-SMT-10} format
\item an SMT solver of choice (we used primarily CVC4 \cite{webcvc4}, and demonstrated interoperability with Z3 \cite{DeMoura:2008:ZES:1792734.1792766}) is applied to the VC
\item if the VC is invalid, a counterexample is obtained from the SMT tool
\item the logical variables of the counterexample are translated back to source code variables and 
text locations; the values of logical variables are expressed in programming language terms
\item the counterexample values and the static ``execution'' path are displayed in the source code editor by hover information and highlighting
\end{itemize}

Using SMTLIB allows us to plug-and-play any SMT solver, with some limitations. SMT solvers still have individual idiosyncrasies that require some tailoring of the VC structure. In addition, there are some valuable features
that are missing from standard SMTLIBv2, such as conversion between Int values and Bit-Vector values,  defining constant arrays, and reasoning about partially ordered sets (such as inheritance hierarchies). SPEEDY requires SMT solvers that produce counterexamples. SPEEDY also requires a combination of a number of SMT logics: Arrays, Uninterpreted Functions and either BitVectors or a combination Int and Real non-linear arithmetic. Most preferable would be a combination
of all of the above, but not all SMT tools support this range of logics. Our demonstration can use either CVC4 or Z3.

We experimented with three techniques for verifying C code and specifications. First we combined the code and specifications into ACSL-annotated C code and checked the result using Frama-C's \cite{webframac} WP plugin (which uses alt-ergo \cite{altergo-web} as an SMT solver). This successfully verified simple procedures and specifications but does not return counterexample information. Second we performed our own conversion of C source code and its specifications to SMTLIB logical assertions, submitting the result directly to CVC4 or Z3 as SMT solvers. In this case we can obtain counterexample information from the solver, displaying it for debugging purposes as described below. A third experiment is to use Frama-C with the value analysis plugin. Of these, the second is most useful for our purposes. However, the task of creating a full translation of C and specifications to a logical encoding is substantial, a task that Frama-C already performs (for part of C, but not C++). Thus we are continuing to investigate means to obtain
counterexample information through Frama-C.

From an IDE perspective, the key point is the ability to debug specifications in source code terms. Early tools, such as Simplify \cite{Detlefs-Nelson-Saxe03} and Esc/Java \cite{Leino-Nelson-Saxe00}, provided counterexample information as an internal dump of the solver state, using many internal variable names. This information was inscrutable even to experts. The workflow above, combined with presentation in the IDE, solves this problem. This technology was adapted for C and SPEEDY from an implementation for OpenJML \cite{Cok-2011-OpenJML} for Java and the Java Modeling Language (JML) \cite{Burdy-etal03,JMLweb}.

\begin{figure}
\includegraphics[width=\textwidth]{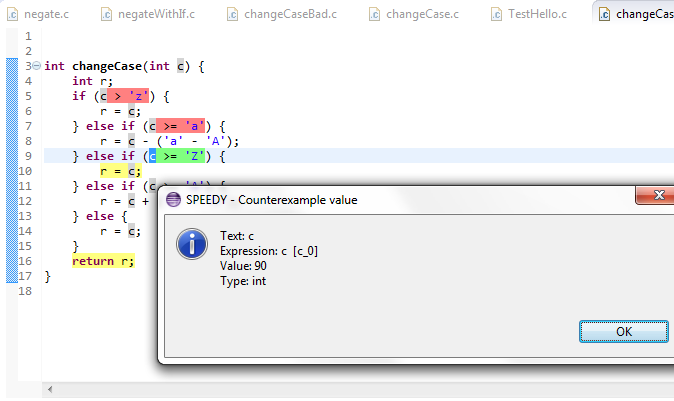}
\caption{SPEEDY GUI showing a counterexample path and variable values}
\label{Fig:counterexample}
\end{figure}

The SPEEDY GUI for debugging specifications is shown in Fig. \ref{Fig:counterexample}. Note that this is {\em static} debugging: the constraint solver seeks an assignment of values to variables that satisfies the object invariants, preconditions, background predicates, and the code itself, but falsifies one or more of the assertions, postconditions or ending invariants of the specifications. 
If found, such a counterexample contains a set of inputs which, if used as the inputs in an actual execution, would trigger the identified assertions. The static debugger then allows the user to 
observe the control flow and variable values along the entire path, moving back and forth along the path at will. 

The figure shows the execution path as highlighted statements -- yellow for executed statements, red
for boolean conditions (e.g., branch conditions and pre or post conditions) that are false, and green for
those that are true. Variable values are shown as hover information or in dialog boxes.

\section{Tools for inferring specifications}
\label{Inference}

SPEEDY incorporates two ways of saving the user effort in generating specifications. One is the set of 
user experience design elements described above; the other is an integrated collection of tools and algorithms that 
directly propose invariants based on program analysis.

SPEEDY's architecture is designed to be open to existing and future specification discovery techniques. In this development phase, we integrated five mechanisms for discovering specifications. These were chosen as the first to be implemented in order to validate that the overall architecture could accommodate a variety of tools.
In the next phase of development we will implement other algorithms from the verification literature.
We have selected algorithms that use a variety of techniques (e.g., static analyses, dynamic analyses, symbolic execution, abstract interpretation, SMT solvers, ...) and that target various kinds of invariants (loop invariants, object invariants, preconditions, postconditions, frame conditions, memory safety, ...).
\begin{itemize}
\item \textbf{Specifications from CodeSonar.} CodeSonar \cite{CodeSonar} is a commercial tool for automatically finding flaws, such as buffer overruns, in a program using static analysis; it does not use specifications. Rather, it performs
a bottom-up program analysis, inferring function summaries for each procedure. The function summaries
include {\em triggers}, which are logical conditions that, if true in the calling context, indicate
that a warning should be issued to the user. Thus the (negations of) triggers serve as preconditions for the 
procedure. SPEEDY imports the triggers from CodeSonar, expressing them in source code terms, 
as preconditions understandable by the user and checkable by other tools.

\item \textbf{Specifications from Daikon.} Daikon \cite{ErnstPGMPTX2007} is a dynamic analysis tool. A front-end tool instruments the target program 
to record the values of various variables at various program points; then the target program is executed, perhaps over many input samples, and the trace data is collected. Daikon analyzes the
trace data to find invariants that are true of all of the traces. SPEEDY then imports those invariants.
Since Daikon's invariants are not necessarily true of all possible
executions (just of the observed executions), these invariants will need human review. Daikon's invariants can be preconditions, 
postconditions, or object invariants.

\item \textbf{Specifications from CodeSurfer.} CodeSurfer \cite{CodeSurfer} is a static analysis tool that produces a wealth of program analysis information about a piece of software. One example is data dependencies. For this prototype, we chose to 
import from CodeSurfer information about the use of global variables by a procedure. This information
can be presented as frame conditions.

\item \textbf{Specifications from Frama-C.} Frama-C \cite{webframac} is also a static analysis tool that produces program
analysis data. In this case, we used Frama-C's value analysis plug-in to obtain information about the
set of values a variable can have at a given program point. This information is translated into invariants; it is also used to attest to the possibility or impossibility of various assertions being
falsified.

\item \textbf{Specifications by symbolic execution.} The above tools are all external tools, performing their own analyses, whose results we integrated into SPEEDY. The last integration demonstrates SPEEDY's
ability to implement and integrate a specific algorithm. For the prototype, we symbolically executed the target procedure along each program path to identify preconditions, postconditions and frame conditions for the procedure. This algorithm can only handle simple procedures (loop-free, not too many branches, simple logic), but it can generate full specifications for many simple procedures, 
saving the user time and effort.

\end{itemize}

A point to note in this section is that it is important for a tool like SPEEDY to incorporate many
kinds of specification discovery techniques: some generate preconditions, some object invariants, some loop invariants; tools will each have their own strengths and weaknesses.

\section{Sources of AST, type, and static analysis information}
\label{ASTs}

Writing an analysis algorithm that generates specifications from source code is helped by
beginning with a representation of already-parsed and type-checked program text. Thus as part of SPEEDY's
design we investigated what such sources were available and usable in our context. We found four:
\begin{itemize}
\item Eclipse CODAN - CODAN is Eclipse's internal Code Analysis framework for C/C++. It is completely integrated 
in Eclipse, providing Java APIs for all the relevant parse tree (AST) and type information. We used it to implement the symbolic execution algorithm described above.
\item Clang \cite{Clang} - Clang's C/C++/Objective C compiler includes its own code analysis framework, intended
for extension by users to provide additional static checks while compiling. Clang is an open-source, commercially-supported tool under active development, so it is readily available for use by a project like SPEEDY. Its interface is C. Clang is based on LLVM and can support analysis of any programming language for which translation to LLVM is implemented.
\item CodeSurfer - GrammaTech's CodeSurfer tool also has an API for AST and program analysis information. It is also commercially released and supported. It has native C and Scheme interfaces, and through SWIG \cite{webSWIG}, many other language bindings.
\item Frama-C - Frama-C  \cite{webframac} is an open-source static analysis environment constructed for program analysis and software
verification. Many algorithms and plug-ins have been implemented on it, and it has been used for 
verifying safety-critical software in the European avionics industry. Its interface is ML.
\end{itemize}
One of the challenges in the next prototyping phase will be selecting among these as the foundation for algorithm implementations; we hope in fact, to be able to implement algorithms in a way that interfaces to 
multiple analysis infrastructures,despite the differing APIs.

\section{Related work}

User interfaces for several other tools have been built using Eclipse: ESC/Java2 \cite{Kiniry-Cok05}, OpenJML \cite{Cok-2011-OpenJML,Cok-2014-OpenJML}, and jSMTLIB \cite{Cok-2011-jSMTLIB} are examples. Similarly Spec\# \cite{SpecSharp2005} and Dafny \cite{Leino:Dafny:LPAR16,Dafny2014} provided such functionality within Visual Studio. Leon \cite{Blanc:2013:OLV:2489837.2489838}, Key \cite{beckert2007verification} and the Why system \cite{Why3} have interfaces of their own. There are also tools primarily focused on proof management, such as PVS\cite{cade92-pvs,OwrShaRusStr01b}, Coq \cite{opac-b1101046}, and Isabelle \cite{Paul94}.

The counterexample debugging technology used here was pioneered in OpenJML \cite{Cok-2011-OpenJML}. OpenJML (for Java) also shares the vision of software verification technology thoroughly integrated into software development
IDEs and into software developers' daily work habits.
The user guidance features 
integrated in SPEEDY (for C) take inspiration from VeriWeb \cite{SchillerE2012} (a web-based research tool for Java).

There are very many papers describing invariant inference algorithms. It is not the point of SPEEDY to develop new such algorithms, though adaptation will certainly be required. Instead, SPEEDY integrates a
number of inference algorithms and techniques, with the goals of enabling easy application and comparison of these techniques. SPEEDY's innovation here is in integrating a number of such techniques into a software development IDE.

\section{Conclusions and Future Work}

At this point SPEEDY is a 6-month-old prototype funded by an initial contract from NASA.
The Phase I work concentrated on resolving identified risks and designing an overall architecture.
The prototype demonstrates all the basic UI concepts described in the paper and integrates a selection of 
invariant inference techniques. 
If the contract is renewed (or other funding is obtained), the prototype will be expanded into
a deployable tool that can serve as a foundation for software verification in C, for studying
invariant inference algorithms, and for evaluating designs for programming and specification productivity.
Thus any conclusions based on research to date are very preliminary, but the work so far supports our key ideas:
\begin{itemize}[noitemsep,nolistsep]
\item Eclipse, the Eclipse CDT, and the CODAN framework make a solid base for an IDE for formal methods tools
\item Integrating specification manipulation features directly in the software development environment streamlines the work of creating correct specifications, even for specification experts.
\item Our architecture for integrating various invariant inference techniques accommodates a wide variety of tool types.
\item Eclipse's extension capabilities are adequate for the user assistance GUI features implemented or planned for SPEEDY.
\item Displaying and interacting with counterexample information in the source code editor is highly effective; this design ported from Java to C without difficulty.
\end{itemize}

\vspace{1em}
One intent of the project is to integrate and compare a variety of techniques; the results of that comparison are in progress and will be reported later. Another goal, however, is to create an effective IDE for working with specifications. Having a variety of techniques, with different idiosyncrasies and perhaps slightly different formal bases may be a disadvantage in creating a unified, coherent IDE and underlying specification manipulation system. The alternative is to create a fully new system or to expand one of the currently available choices. This alternative would be much more costly (in development labor). We will understand the drawbacks of the integrative approach better after the next phase of research.

The following additional work is planned:
\begin{itemize}[noitemsep,nolistsep]
\item Study and integration of additional invariant inference algorithms
\item Integration with other SMT-based constraint solvers
\item Implementing additional user-guidance techniques for advising about specification generation, review, correction, and debugging
\item Addition of other UI features, in particular
\begin{itemize}[noitemsep,nolistsep]
\item integration with Eclipse's refactoring and searching mechanisms
\item quick fix suggestions for erroneous specifications (or code)
\item extending syntax highlighting to specifications
\end{itemize}
\item Scaling up the capability and set of application experiments to larger sets of code
\item Comparing and evaluating invariant inference algorithms
\end{itemize}

\vspace{1em}
Our work on C/C++ so far has used ACSL and the Frama-C tool. However, both of these only address a portion of ANSI-C (because they were developed to verify safety-critical avionics, which uses a `safe' subset of ANSI-C). There is no full-fledged Behavioral Interface Specification Language for C++, similar to JML, Spec\#, ACSL, or VCC. Such a language would need to combine the object-oriented features of JML and Spec\# with the detailed memory buffer features contained in VCC and ACSL. Developing a C++ specification language is a significant task best addressed by the specification and verification research community as a whole.

\section{Acknowledgments} 
The majority of this work was performed by the authors at GrammaTech, Inc., under contract to NASA (SBIR Phase I contract \#NNX13CL55P). Some of the results described here are based upon work supported by the National Science Foundation under Grant No. ACI-1314674. Any opinions, findings, and conclusions or recommendations expressed in this material are those of the authors and do not necessarily reflect the views of the National Science Foundation.
\section{Bibliography}

\bibliographystyle{eptcs}

\end{document}